\documentclass[preprint,pre,showpacs]{revtex4}
\usepackage{amscd}
\usepackage{amsfonts}
\usepackage{amsmath}
\usepackage{amssymb}
\usepackage{undertilde}
\usepackage{color}
\usepackage{graphicx}
\usepackage{graphics}
\usepackage{hyperref}
\usepackage{wrapfig}
\usepackage[T1]{fontenc}
\usepackage[latin1]{inputenc}
\usepackage{gensymb}
\def\>{\rangle}
\def\<{\langle}
\def\n{\nonumber}

\newcommand{\id}{\openone}
\begin{document}
\title{\textit{Ab initio} relaxation times and time-dependent Hamiltonians within the steepest-entropy-ascent quantum thermodynamic framework}
\author{Ilki Kim$^{1}$}
\email{hannibal.ikim@gmail.com}
\author{Michael R. von Spakovsky$^{2}$}
\affiliation{$^{1}$Center for Energy Research and Technology, North
Carolina A$\&$T State University, Greensboro, NC 27411\\
$^{2}$Department of Mechanical Engineering, Virginia Tech, Blacksburg, VA 24061}
\date{\today}
\begin{abstract}
Quantum systems driven by time-dependent
Hamiltonians are considered here within the framework of steepest-entropy-ascent quantum
thermodynamics (SEAQT) and used to study the thermodynamic characteristics of such
systems. In doing so, a generalization of the SEAQT framework valid
for all such systems is provided, leading to the development of an \textit{ab initio} physically relevant expression for the intra-relaxation time, an important element of this framework and one that had as of yet not been uniquely determined as an integral part of the theory. The resulting expression for the relaxation time is valid as well for time-independent Hamiltonians as a special case and makes the description provided by the SEAQT framework more robust at the fundamental level. In addition, the
SEAQT framework is used to help resolve a fundamental issue of thermodynamics in the
quantum domain, namely, that concerning the unique definition of {\em process-dependent} work and heat functions. The developments presented lead to the conclusion that
this framework is not just an alternative approach to thermodynamics in the quantum domain but instead one that uniquely sheds new light on various fundamental but as of yet not completely resolved questions of thermodynamics.
\end{abstract}
\maketitle
%
\section{Introduction}\label{sec:introduction}
The last three decades have seen experimental evidence (e.g.,
\cite{BRUN1,TURCH1,BOLL,WALS,CHUPP,MAJU,BENAT,LISI,KLAPD,HOOP,BATAL})
emerge at atomistic scales, which suggests the existence of
irreversible changes even at these scales. Whether or not these
changes are related to the measurement axiom of quantum mechanics
(QM), the so-called ``collapse of the wave function'', i.e., an
abrupt collapse leading to irreversible change, or to something else
entirely different is still a matter of debate. What is clear is
that the collapse of the wave function postulate has drawn
significant criticism
\cite{EVER1,EVER2,MARG1,MARG2,PARK1,PARK2,GUERL,GLEY,SAYR} and has
led to an interpretation which replaces the abrupt collapse by a
more gentle differentiable dynamical evolution. The result has been
two theories, i.e., that of quantum open systems (QOS)
\cite{KRAUS,LIND,GORI,BREUER} and that of typicality
\cite{GEMM1,GOLD1,LUBK,GOLD2,REIM,POPE} from which it is said that
the Second Law of thermodynamics emerges. The former, which is a
special case of the latter, relies on a partition between the
primary system and the environment (e.g., the measuring device) and
the total evolution in state is assumed to be unitary (i.e., linear)
and generated by the Hamiltonian of the system-environment
composite.

An alternative to an assumed collapse whether abrupt or more gradual
is a possibly meaningful, nonlinear dynamics, which results when the
postulates of QM are complemented by the Second Law, which, instead
of emerging from QM, supplements it. In such an approach, the
evolution of state can occur non-unitarily consistent with both the
postulates of QM and thermodynamics. One such approach is that of
intrinsic quantum thermodynamics (IQT)
\cite{HATS1,HATS2,HATS3,HATS4,BERET1,SIMM,BERET2,BERET3,BERET4,BERET5}
and its mathematical framework steepest-entropy-ascent quantum
thermodynamics (SEAQT)
\cite{BERET6,BERET7,BERET8,BERET9,BERET10,LI1,LI2,LI3,CANO1,SMITH,CANO2,LI4,LI5,BERET11,LI6}.
It is this approach and the ones described above that are
representative of the contrasting views of the origins of
irreversible changes that form the basis of the field of
\textit{quantum thermodynamics} \cite{GEMM1,VINJ,VONS1}, which has
developed over the last four decades and has grown exponentially in
the last decade and a half. In fact, the term \textit{quantum
thermodynamics} was first coined by Beretta \textit{et al}.
\cite{BERET1,SIMM,BERET2,BERET3,BERET4} in the early 1980s with the
publication of the dynamical aspects of IQT.

It is the mathematical framework of this latter theory, i.e., SEAQT,
which is the basis of the developments presented here. Differing
from other known approaches, the SEAQT framework results from a
unified treatment of quantum mechanics and thermodynamics at a
single level of description based on a generalized scheme of quantal
dynamics in which the standard unitary dynamics governed by a given
Hamiltonian is supplemented by a intra-dissipative (non-unitary)
dynamics obtained from the requirement of maximum entropy production
at every single instant of time. Remarkably enough, this enables the
Second Law of thermodynamics to appear straightforwardly at a
fundamental level of description (cf. for a contrasting view based
on a quantum Maxwell demon, see \cite{LEB16,LES16,LLO97}). As such,
the SEAQT framework, which has been shown to encompass all of the
well-known classical and quantum non-equilibrium frameworks
\cite{BERET10} and is applicable even far from equilibrium, provides
a conceptually consistent and mathematically and relatively compact
framework for systematically analyzing non-equilibrium processes at
any spatial and temporal scale. This first-principles,
thermodynamic-ensemble-based approach has recently been extended via
the concept of hypoequilibrium state and a corresponding set of
intensive properties \cite{LI1} to provide the global features of
the microscopic description as well as that of the nonequilibrium
evolution of state of a system when combined with a set of
nonequilibrium extensive properties. In contrast to the definitions
of other nonequilibrium thermodynamic approaches, the SEAQT
intensive property definitions are fundamental as opposed to
phenomenological, are applicable to all nonequilibrium states, and
enable the generalization of the equilibrium and near-equilibrium
description (e.g., the Gibb's relation, the Clausius inequality, the
Onsager relations, and the quadratic dissipation potential) to the
far-from-equilibrium realm. In addition, reduced computational
burdens make the study of physically complex non-equilibrium
phenomena at micro-scales possible where otherwise they may not be
given the much heavier computational burdens associated with
conventional approaches based purely on mechanics (i.e., quantum or
classical) and/or stochastics (e.g., ensemble Monte Carlo). This
framework can also facilitate the development of micro-scale
analytical expressions, and its extension from the quantum to the
classical regime is accomplished without resort to any extra (semi)
classical approximations and manipulations, which are normally
non-uniquely made. As a consequence, this approach provides a robust
platform for exploring the thermodynamics of the quantal-classical
transition regime and for affecting the scale-up of systems
consisting of a few qubits to those of much greater extent, doing so
with a single unified multi-scale thermodynamic picture of the
kinematics and dynamics involved.

Both reactive and non-reactive quantum and classical systems have
been investigated successfully using SEAQT
\cite{LI1,LI2,LI3,CANO1,SMITH,CANO2,LI4,LI5,BERET11,LI6} and some
validations with experiment have been made \cite{CANO1,SMITH}. It
has furthermore been shown that not only does the equation of motion
of SEAQT predict the unique thermodynamic path, which the system
takes, \cite{BERET2,BERET3} but that the kinetics of this path
(i.e., movement along it) and its dynamics (i.e., the time it takes
for this movement) can be treated separately \cite{LI2}. Physically,
this means that the system follows the same trajectory in state
space regardless of the relaxation time $\tau$ chosen for the
equation of motion. Whether a constant or a functional of the
density operator $\hat{\rho}$ upon which the equation of motion is
based, the dynamics of the process and, as a consequence a value for
$\tau$, is determined via experiment \cite{CANO1,SMITH,BERET11} or,
for example, kinetic theory \cite{LI2,LI4,LI6}. What has been
missing to date is $\tau$ as a functional of $\hat{\rho}$. Although
Beretta \cite{BERET9} by analogy provides a lower limit for the
relaxation time relative to the time-energy Heisenberg uncertainty
principle, this limit does not in general, as has been shown in
\cite{CANO1,SMITH,CANO2,LI4,LI6}, provide a practical value for
$\tau$. The purpose of the present paper is to provide such a
functional and as a consequence a generalization of the SEAQT
framework both for time-independent and time-dependent Hamiltonians.
This development appears in Section \ref{sec:sec4}. An added benefit
of this development is that since the SEAQT framework inherently
satisfies all the laws of mechanics and thermodynamics, generalized
concepts for process-dependent heat and work transfers and
process-independent internal energy changes in the quantum domain
are provided. This appears in Section \ref{sec:sec3} and results in
the First Law of thermodynamics and its resulting energy balance
being uniquely well defined in the quantum domain, remarkably enough
with the help of the Second Law, which the SEAQT framework embodies.
We begin in Section \ref{sec:sec2} with an introduction to the SEAQT
equation of motion and the limits placed on the relaxation times
associated with the Hamiltonian and dissipation terms of this
equation.

%
\section{Relaxation Time Limits and the SEAQT Equation of Motion}\label{sec:sec2}
%
In the SEAQT framework for a single isolated system with a
time-independent Hamiltonian, the time evolution of a density
operator is given by \cite{BERET9}
\begin{equation}\label{eq:von-neumann1}
    \frac{\partial \hat{\rho}}{\partial t}\, =\,
    \frac{i}{\hbar}\,[\hat{\rho}, \hat{H}] + \hat{\mathbf{D}}_1(\hat{\rho}, \hat{H},
    \{\hat{N}_j\})\,,
\end{equation}
where $\hat{H}$ is the Hamiltonian, $\hat{\rho}$ the density or
state operator, and  $\hat{N}_j$  the $j$th particle number operator
(or magnetic moment or other operator representing additional
generators of the motion if any). The first term on the right-hand
side governs the reversible dynamics conserving both the energy and
entropy (the so-called von-Neumann term), while the second term
describes the energy-conserving but internally entropy-generating
and, thus, irreversible dynamics and is given by
\begin{align}
    \hat{\mathbf{D}}_1(\hat{\rho}, \hat{H}, \{\hat{N}_j\})\; &=\;
    -\frac{1}{2 \tau}
    \left[\sqrt{\hat{\rho}}\,\hat{D} + \hat{D}^{\dagger} \sqrt{\hat{\rho}}\right]\tag{\ref{eq:von-neumann1}a}\label{eq:dissipation1}\\
    \hat{D}\; &=\; \left.\sqrt{\hat{\rho}}\,\hat{B}\,\ln(\hat{\rho})\right|_{\perp\,{\mathcal
    L}\{\sqrt{\hat{\rho}}\,\id,\,\sqrt{\hat{\rho}}\,\hat{H},\,[\sqrt{\hat{\rho}}\,\hat{N}_j]\}}\,.\tag{\ref{eq:von-neumann1}b}\label{eq:dissipation2}
\end{align}
In standard quantum mechanics, which neglects the entropy-generating
term $\hat{\mathbf{D}}_1$, Eq. (\ref{eq:von-neumann1}) obviously
reduces to the well-known von-Neumann equation, giving rise to the
unitary time evolution $\hat{\rho}(t) =
\hat{U}\,\hat{\rho}(t_0)\,\hat{U}^{\dagger}$ with $\hat{U}(t, t_0) =
e^{-i (t-t_0) \hat{H}/\hbar}$. In
(\ref{eq:dissipation1})-(\ref{eq:dissipation2}) the intra-relaxation
time $\tau = \tau(\hat{\rho})$ is a positive functional of
$\hat{\rho}$, but has not uniquely been determined as of yet
\cite{BERET9}. The idempotent operator $\hat{B}$ is introduced which
assigns unity for each non-zero eigenvalue of $\hat{\rho}$, while
zero for each vanishing eigenvalue of $\hat{\rho}$, thus, ensuring
that the entropy operator $\hat{S} = -k_{\mbox{\tiny B}}
\hat{B}\,\ln(\hat{\rho})$ is well-defined even when some eigenvalues
of $\hat{\rho}$ vanish. By construction, the operator $\hat{D}$ is
the component of $\hat{\mathbf{D}}_1$ perpendicular to the linear
manifold ${\mathcal L}$ spanned by a set of operators
$\{\sqrt{\hat{\rho}}\,\id,\,\sqrt{\hat{\rho}}\,\hat{H},\,[\sqrt{\hat{\rho}}\,\hat{N}_j]\}$.
The operator $\hat{\mathbf{D}}_1$ is then interpreted as driving the
density operator $\hat{\rho}(t)$ at every instant of time in the
direction of steepest entropy ascent ($ds/dt|_{\text{\tiny max}} >
0$ with the entropy $s = -k_{\text{\tiny
B}}\,\text{Tr}\{\hat{\rho}\,\ln\hat{\rho}\}$) relative to manifold
specified by the time invariants $\{{\mathcal U} =
\text{Tr}(\hat{\rho} \hat{H}), [{\mathcal N}_j =
\text{Tr}(\hat{\rho} \hat{N}_j)]\}$. Here ${\mathcal U}$ is the
internal energy of the system and ${{\mathcal N}_j}$ the number of
particles of the $j$th constituent.

Eq. (\ref{eq:von-neumann1}) can be rewritten in the alternative form
\cite{BERET9}
\begin{equation}\label{eq:total_E_1}
    \frac{\partial \hat{\rho}}{\partial t} = \sqrt{\hat{\rho}}\,\hat{E}(t) + \hat{E}^{\dagger}(t)\,\sqrt{\hat{\rho}}\,,
\end{equation}
which will be used below. Here the decomposition $\hat{E} =
\hat{E}_H + \hat{E}_D$ is composed of the von-Neumann part $\hat{E}_H =
(i/\hbar) \sqrt{\hat{\rho}}\,\{\hat{H} + c(\hat{\rho}) \id\}$
corresponding to standard quantum mechanics with $c(\hat{\rho})
\in {\mathbb R}$ being an arbitrary functional of $\hat{\rho}$
and the entropy-generating part $\hat{E}_D = -\hat{D}/(2 \tau)$.
As a consequence, the total dynamics given in (\ref{eq:von-neumann1})
is non-unitary as long as the initial state $\hat{\rho}(t_0)$
is in form of a mixed state. For any {\em any} pure state $\hat{\rho}(t_0) = |\psi(t_0)\>\<\psi(t_0)|$, the dynamics becomes unitary with $\hat{E} \to \hat{E}_H$. In this case, the operator $\hat{D}$ identically vanishes at every instant so that no entropy is generated during the time evolution. It is
also straightforward to show that since $\sqrt{\hat{\rho}}$ is perpendicular to both components of $\hat{E}$,
\begin{equation}\label{eq:density -op-conserved_1}
    0 = \text{Tr}\left(\frac{\partial \hat{\rho}}{\partial t}\right) =
    \text{Tr}\left\{(\hat{E} + \hat{E}^{\dagger})\,\sqrt{\hat{\rho}}\right\} =
    2\,\left(\hat{E}|\sqrt{\hat{\rho}}\right)\,,
\end{equation}
where the inner product $(\hat{F}|\hat{G}) =
\{\text{Tr}(\hat{F}^{\dagger} \hat{G} + \hat{G}^{\dagger}
\hat{F})\}/2$ in symmetrized form is defined in the space
$L({\mathcal H})$ of linear operators on the Hilbert space
${\mathcal H}$.

The operator $\hat{E}_H$ can directly be related to the time-energy
uncertainty relation by first setting the real number $c(\hat{\rho})
= -{\mathcal U}$ such that $\hbar^2 (\hat{E}_H|\hat{E}_H) =
(\sqrt{\hat{\rho}}\,\Delta \hat{H}|\sqrt{\hat{\rho}}\,\Delta
\hat{H}) = \sigma^2_{\scriptscriptstyle H}$ with $\Delta \hat{H} =
\hat{H} - {\mathcal U} \id$ the deviation operator of $\hat{H}$ and
$\sigma_{\scriptscriptstyle H}$ the standard deviation relative to
$\hat{H}$ \cite{BERET9}. It then turns out that
$\tau^{-2}_{\scriptscriptstyle H} := 4\,(\hat{E}_H|\hat{E}_H) \geq
\tau^{-2}_{\scriptscriptstyle A}$ with the help of the uncertainty
relation $\sigma_{\scriptscriptstyle H}\,\tau_{\scriptscriptstyle A}
\geq \hbar/2$ where the characteristic time
$\tau_{\scriptscriptstyle A}$ for a given observable $\hat{A}$ (not
explicitly time-dependent) may be interpreted as the amount of time
it takes the expectation value of $\hat{A}$ to change by one
standard deviation $\sigma_{\scriptscriptstyle A} = |d
\<\hat{A}\>/dt|\,\tau_{\scriptscriptstyle A}$ \cite{MES76,GRI05}.
Accordingly, the time $\tau_{\scriptscriptstyle H}$, which results
from the time-energy uncertainty relation with strict equality, is
simply chosen above as the minimum value of the characteristic
times, the $\tau_{\scriptscriptstyle A}$'s, for {\em all} possible
observables, i.e., $\hat{A}$'s. Analogously, in \cite{BERET9}, it is
assumed that the entropy-generating part $\hat{E}_D$ also satisfies
the uncertainty relation, which renders the corresponding
characteristic time $\tau_{\scriptscriptstyle D} =
\tau\,(\hat{D}|\hat{D})^{-1/2}$ for which a value is found from the
uncertainty equality \cite{BERET9}. This minimum value
$(\tau_{\scriptscriptstyle D})_{\text{\tiny min}}$ provides the
intra-relaxation time $\tau$ in question with its minimum value
$\tau_{\text{\tiny min}}$.

However, this value $\tau_{\text{\tiny min}}$ has been shown to be
significantly too small for generic experimental values of the
relaxation time $\tau$, and so the substitution of
$\tau_{\text{\tiny min}}$ into (\ref{eq:dissipation1}) cannot be
supported by the experimental data. Also theoretically, it has been
verified that a minimum-uncertainty state must be a pure state
\cite{STO72,BAL98}. In other words, the intra-relaxation time
$\tau(\hat{\rho})$ for a mixed state $\hat{\rho}$ is required to be
fundamentally greater than its minimum-uncertainty value. As a
result, it is not physically consistent to impose the value
$\tau_{\text{\tiny min}}$ upon the time evolution given in
(\ref{eq:total_E_1}) for a generic mixed state. To address this, we
introduce a different approach below for the determination of the
intra-relaxation time, which is more physically relevant.

%
\section{Generalization of the SEAQT Equation of Motion for a Time-dependent Hamiltonian}\label{sec:sec3}
%
Now to generalize Eq. (\ref{eq:total_E_1}) for the case of a
time-dependent Hamiltonian $\hat{H}(t)$, the corresponding
von-Neumann part $\hat{E}_H$ is first determined. From the
von-Neumann equation valid also for this case, it easily follows
that $\hat{E}_H = (i/\hbar) \sqrt{\hat{\rho}}\,\Delta\hat{H}(t)$,
thus leading to
\begin{equation}\label{eq:time-dpt-inner-products_1}
    (\hat{E}_H|\sqrt{\hat{\rho}}\,\id) = 0\;\; ;\;\;
    (\hat{E}_H|\sqrt{\hat{\rho}}\,\hat{H}) = (\hat{E}_H|\sqrt{\hat{\rho}}\,\Delta\hat{H}) =
    0\,.
\end{equation}
Also note that
$(\sqrt{\hat{\rho}}\,\id|\sqrt{\hat{\rho}}\,\Delta\hat{H}) = 0$,
i.e., these two operators are perpendicular to each other. It is
also true that the energy-time uncertainty relation with the
minimum-uncertainty equality holds true for this case (cf.
\cite{KIM17}). For purposes of comparison below, the unitary
operator of time-evolution $\hat{U}(t) = \hat{T} e^{-i \int_0^t
\hat{H}(t)/\hbar}$ of standard quantum mechanics obtained from the
von-Neumann equation is briefly discussed. Here, the operator
$\hat{T}$ denotes the time-ordering. In most of cases, it is a
highly non-trivial exercise to derive a closed form expression for
this operator. Nonetheless, it is instructive to transform this
time-ordered form to an ordinary exponential form as in the case of
a time-independent Hamiltonian. Thus, the exponential operator
identity is applied such that \cite{LAM98}
\begin{equation}\label{eq:exponential_identity1}
    \hat{T}\,\exp\left\{\int_0^t d\tau\,\hat{\mathcal{B}}_t(\tau)\right\}\, =\,
    \exp\left\{\sum_{n=1}^{\infty} \hat{K}_n(t)\right\}\,,
\end{equation}
where some of the lower-order terms are explicitly given by
\begin{align}
    \hat{K}_1(t)\, =\, \hat{C}_1(t) \;\;\;&;\;\;\; \hat{K}_2(t)\, =\, \frac{1}{2}\,\hat{C}_2(t)\tag{\ref{eq:exponential_identity1}a}\label{eq:exponential_identity2}\n\\
    \hat{K}_3(t)\, =\, \frac{1}{3}\,\hat{C}_3(t) + \frac{1}{12}\,\left[\hat{C}_2(t), \hat{C}_1(t)\right] \;\;\;&;\;\;\;
    \hat{K}_4(t)\, =\, \frac{1}{4}\,\hat{C}_4(t) + \frac{1}{12}\,\left[\hat{C}_3(t),
    \hat{C}_1(t)\right]\,.\n
\end{align}
Here the commutators are written as
\begin{align}
    \hat{C}_n(t)\, =\, \int_0^t d\tau_1 \int_0^{\tau_1} d\tau_2 \cdots \int_0^{\tau_{n-1}} d\tau_n\,
    \left[\hat{\mathcal{B}}_t(\tau_1), \left[\hat{\mathcal{B}}_t(\tau_2),
    \left[\cdots, \left[\hat{\mathcal{B}}_t(\tau_{n-1}), \hat{\mathcal{B}}_t(\tau_{n})\right] \cdots\right]\right]\right]\tag{\ref{eq:exponential_identity1}b}\,,\label{eq:exponential_identity3}
\end{align}
where $\hat{C}_1(t) = \int_0^t d\tau\,\hat{\mathcal{B}}_t(\tau)$
with $\hat{\mathcal{B}}_t(\tau) = -(i/\hbar) \hat{H}(\tau)$. In
fact, the operators $\hat{K}_n(t)$ for all $n$ can be evaluated
exactly. As an example of the time evolution in closed form,
consider the two-level system given by $\hat{H}_0(t) =
(\hbar\omega/2) \{\hat{\sigma}_z + a(t)\,\hat{\sigma}_x\}$ where the
$\hat{\sigma}_j$'s denote Pauli matrices and a dimensionless
quantity $a(t) \in {\mathbb R}$ is periodic in time $t$. The
system's time evolution is then explicitly given by $\hat{U}_0(t) =
\hat{U}_y\,\utilde{\hat{U}}(t)\,\hat{U}_y^{\dagger}$ with $\hat{U}_y
= e^{(i\pi/4)\,\hat{\sigma}_y}$ and the $(2 \times 2)$ matrix
$\utilde{\hat{U}}(t)$ by \cite{ARA84,HAE98,BAR00}
\begin{equation}\label{eq:unitary-evolution1}
    \utilde{\hat{U}}(t) = \left(\begin{array}{ccc}
                                    R(t)\,\{1 + i\,g_0\,S(t)\}&&-i\,\epsilon\,R(t)\,S(t)\\
                                    -i\,\epsilon\,\bar{R}(t)\,\bar{S}(t)&&\bar{R}(t)\,\{1 - i\,\bar{g}_0\,\bar{S}(t)\}
                                \end{array}\right)\,,
\end{equation}
where $\epsilon = \hbar\omega/2$, and $\bar{R}$, $\bar{S}$ and
$\bar{g}_0$ denote the complex conjugates of $R$, $S$ and $g_0$,
respectively. Here $R(t) = \exp[-i \int_0^t\,\{f(t') +
g(t')\}\,dt']$, and $S(t) = \int_0^t\,\{R(t')\}^{-2}\,dt'$ where
$f(t) = -(\hbar\omega/2)\,a(t)$ and $g(t)$, with $g_0 = g(0)$, is a
particular solution to the generalized Riccati equation
$\partial_t\,g(t) - i g^2(t) - 2 i f(t)\,g(t) + i \epsilon^2 = 0$
\cite{GRA07}.

Next, the corresponding entropy-generating part $\hat{E}_D$ of Eq.
(\ref{eq:total_E_1}) is determined. To begin with, the energy
balance of the First Law of thermodynamics is written as (see, e.g.,
\cite{ABE12})
\begin{equation}\label{eq:1st-law1}
    d{\mathcal U} = \sum_n {\mathcal E}_n\,dp_n + \sum_n p_n\,d{\mathcal E}_n\,,
\end{equation}
where the ${\mathcal E}_n$'s are the eigenenergies and the $p_n$'s
their respective probabilities. This balance provides a condition
required for determining the direction of $\hat{E}_D$. The first
term on the right is interpreted as the heat input $\delta
Q_{\text{in}}$ from the environment and the second as the work
$\delta W_{\text{in}}$ performed on the system (cf. see
\cite{VIL08,VIL11,JAR08} for a discussion of the work for classical
systems). Now, consider the case of a time-independent Hamiltonian.
Accordingly, with no work input ($\delta W_{\text{in}} = 0$), the
balance reduces to $d{\mathcal U} = \delta Q_{\text{in}}$, and it is
easily be shown with the help of (\ref{eq:total_E_1}) that
\begin{equation}\label{eq:du_1}
    d{\mathcal U}/dt = \text{Tr}\{\hat{H}\,(d\hat{\rho}/dt)\} = 2 \left(\hat{E}\right.|\left.\sqrt{\hat{\rho}}
    \hat{H}\right)\,.
\end{equation}
Therefore, for an isolated system with no heat exchange ($\delta
Q_{\text{in}} = 0$), $(\hat{E}|\sqrt{\hat{\rho}} \hat{H}) =
0$, which means that the two operators $\hat{E}$ and $\sqrt{\hat{\rho}}
\hat{H}$ are perpendicular to each other. Subsequently, it is also
straightforward to show that $(\hat{E}_H|\sqrt{\hat{\rho}} \hat{H})
= 0$ so that it follows that $(\hat{E}_D|\sqrt{\hat{\rho}} \hat{H})
= 0$ as well. Therefore, the invariance of ${\mathcal U}$ may simply be seen
as resulting from the energy balance in a system with no heat nor work
input. Likewise, for additional non-Hamiltonian
invariants (if any),
\begin{equation}\label{eq:dG_1}
    d{\mathcal N}_j/dt = \text{Tr}\{\hat{N}_j\,(d\hat{\rho}/dt)\} =
    2 \left(\hat{E}\right.|\left.\sqrt{\hat{\rho}}
    \hat{N}_j\right)\,,
\end{equation}
and $d{\mathcal N}_j/dt$ also vanishes. With
$(\hat{E}_H|\sqrt{\hat{\rho}} \hat{N}_j) = 0$, this results in
$(\hat{E}_D|\sqrt{\hat{\rho}} \hat{N}_j) = 0$. Thus, it is seen that
all invariants $\{{\mathcal U}, [{\mathcal N}_j]\}$ uniquely
determine the direction of $\hat{E}$.

Next, a similar scenario is developed for a system with no heat
input but non-zero work input. For this case, the internal energy is
no longer a time-invariant. In fact, it is assumed that there is no
invariant available to the Hamiltonian system given by $\hat{H}(t)$.
The quantity $\text{Tr}(\hat{H} d\hat{\rho})$, as given in
(\ref{eq:du_1}), can then no longer be interpreted as $\delta
Q_{\text{in}}$. To illustrate this, the aforementioned system
$\hat{H}_0(t)$ is now considered. Its instantaneous eigenvalues and
eigenvectors are explicitly given by
\begin{subequations}
\begin{eqnarray}
   {\mathcal E}_1(t) = (\hbar \omega/2) \sqrt{1 + a^2(t)}\;\; &;&\;\; |1(t)\> = {\mathcal N}_+ \left[a(t) |+\> + \{\sqrt{1 + a^2(t)} - 1\} |-\>\right]\label{eq:eigenenergy1}\\
   {\mathcal E}_2(t) = -(\hbar \omega/2) \sqrt{1 + a^2(t)}\;\; &;&\;\; |2(t)\> = {\mathcal N}_- \left[a(t) |+\> - \{\sqrt{1 + a^2(t)} + 1\} |-\>\right]\,.\label{eq:eigenenergy2}
\end{eqnarray}
\end{subequations}
Here the (time-dependent) normalizing numbers are given by ${\mathcal
N}_{\pm} = [2\,\{1 + a^2(t) \mp \sqrt{1 + a^2(t)}\}]^{-1/2}$ with
the signs $+/-$ in accordance with their order on both sides.
The internal energy is then shown to be ${\mathcal U}_0(t) =
\text{Tr}(\hat{H}_0 \hat{\rho}) = \hbar\omega\,[\rho_{11} +
a(t)\,\{\text{Re}(\rho_{12})\} - 1/2]$ where the symbol $\rho_{jk}$
denotes the $(j,k)$-th component of a $(2 \times 2)$-Hermitian
matrix $\hat{\rho}$ and $\text{Re}(\rho_{12})$ is the real part of $\rho_{12}$.
This easily yields $\text{Tr}(\hat{H}_0 d\hat{\rho}) =
\hbar\omega\,[d\rho_{11} + a(t)\,d\{\text{Re}(\rho_{12})\}]$, while
$\text{Tr}(\hat{\rho}\,d\hat{H}_0) = \hbar\omega
\{\text{Re}(\rho_{12})\} da$. In contrast, by expressing
$\hat{\rho}$ within the instantaneous eigenbasis $\{|1(t)\>,
|2(t)\>\}$ of $\hat{H}_0(t)$, its
diagonal elements $p_1 = 1/2 + [\rho_{11} + a
\{\text{Re}(\rho_{12})\} - 1/2]\,(1 + a^2)^{-1/2}$ and $p_2 = 1 -
p_1$ can straightforwardly be obtained. This gives
\begin{subequations}
\begin{eqnarray}
    \delta Q_{\text{in}} &=& \sum_{n} {\mathcal E}_n\,dp_n\, =\, \hbar\omega
    \left[d\rho_{11} + d\{a\,\text{Re}(\rho_{12})\} - \frac{a\,\{\rho_{11} + a\,\text{Re}(\rho_{12}) - 1/2\}\,da}{1 + a^2}\right]\label{eq:heat-explicit-H_0}\\
    \delta W_{\text{in}} &=& \sum_n p_n\,d{\mathcal E}_n\, =\,
    \hbar\omega \left[\frac{a\,\{\rho_{11} + a\,\text{Re}(\rho_{12}) - 1/2\}\,da}{1 + a^2}\right]\,.
\end{eqnarray}
\end{subequations}
In this case, it is seen that $\delta Q_{\text{in}} \ne \text{Tr}(\hat{H}_0
d\hat{\rho})$ and $\delta W_{\text{in}} \ne
\text{Tr}(\hat{\rho}\,d\hat{H}_0)$. Thus, the association of $\delta Q_{\text{in}}$ with $\text{Tr}(\hat{H}_0
d\hat{\rho})$ and $\delta W_{\text{in}}$ with $\text{Tr}(\hat{\rho}\,d\hat{H}_0)$ as is routinely done in the literature (cf. \cite{VONS1,GEMM1}) is not warranted for the case of a time-dependent Hamiltonian.

%
\section{Formal Development of the Relaxation Time Functional}\label{sec:sec4}
%
The previous generalization is now discuss more systematically. To do so, consider
\begin{equation}\label{eq:time-dpt-hamiltonian1}
    \delta \lambda(t) := \text{Tr}\{\hat{H}(t)\,d\hat{\rho}\} = \sum_n {\mathcal E}_n\,\<n|(d\hat{\rho})|n\>\,,
\end{equation}
expressed in terms of the instantaneous eigenvectors $\{|n\>\}$ of
$\hat{H}(t)$. From the identity that $\<n|(d\hat{\rho})|n\> =
d(\<n|\hat{\rho}|n\>) - (d\<n|)\hat{\rho}|n\> - \<n|\hat{\rho}
(d|n\>)$ with $d(\<n|\hat{\rho}|n\>) = dp_n$, it is easily seen that
$\delta \lambda \ne \delta Q_{\text{in}} = \sum_n {\mathcal E}_n
dp_n$ for $\hat{H}(t)$ whereas $\delta \lambda = \delta
Q_{\text{in}}$ for its time-independent counterpart. Thus, for the
case of $\delta Q_{\text{in}} = 0$ and a time-dependent Hamiltonian,
$\delta\lambda \ne 0$ always. Based on Eq. (\ref{eq:du_1}), this
leads to the conclusion that $\hat{E}$ is not perpendicular to
$\sqrt{\hat{\rho}}\,\hat{H}(t)$, which means that the procedure
following Eq. (\ref{eq:du_1}) above for determining the direction of
$\hat{E}$ cannot be employed. However, as seen from
(\ref{eq:time-dpt-inner-products_1}), the von-Neumann part
$\hat{E}_H$ remains perpendicular to
$\sqrt{\hat{\rho}}\,\hat{H}(t)$; and as a consequence, without the
intra-entropy-generation provided by the SEAQT framework (i.e., with
$\hat{E} \to \hat{E}_H$), one must conclude that
$\text{Tr}\{\hat{H}(t)\,(d\hat{\rho}/dt)\} =
2\,(\hat{E}_H|\sqrt{\hat{\rho}}\,\hat{H}) = 0$ [cf.
(\ref{eq:time-dpt-hamiltonian1})], which necessarily contradicts
$\delta\lambda \ne 0$ or $\delta Q_{\text{in}} = 0$. This is a
fundamental conceptual problem within the thermodynamics embedded in
the scheme of standard quantum mechanics. Furthermore, the
entropy-generating part $\hat{E}_D$ cannot be perpendicular to
$\sqrt{\hat{\rho}}\,\hat{H}(t)$ for the case of a time-dependent
Hamiltonian or else $\delta \lambda = 0$, resulting in $\delta
Q_{\text{in}} \ne 0$ which again is a contradiction.

To resolve this conceptual inconsistency and as a result develop a
consistent thermodynamics of the quantum domain, the
intra-entropy-generation available in SEAQT is used to uniquely
determine the direction of $\hat{E}_D$ and as a consequence that of
$\hat{E}$. To that end, it is again assumed that $\delta
Q_{\text{in}} = 0$, i.e.,
\begin{equation}\label{eq:no-heat_0}
    \sum_n \left(\frac{d}{dt} \<n|\hat{\rho}|n\>\right)\,{\mathcal E}_n \stackrel{!}{=}
    0\,,
\end{equation}
so that from the energy balance, $d{\mathcal U} = \delta
W_{\text{in}}$.  Eq. (\ref{eq:no-heat_0}) is subsequently rewritten
as
\begin{align}
    \sum_n \left\<n|(d\hat{\rho}/dt)|n\right\>\,{\mathcal E}_n = - 2 \sum_n
    \text{Re}\left\{\<n|\hat{\rho}\,(d/dt)|n\>\right\}\,{\mathcal E}_n\,.\tag{\ref{eq:no-heat_0}a}\label{eq:no-heat_1}
\end{align}
The left-hand side is nothing else than
$\text{Tr}\{\hat{H}(t)\,(d\hat{\rho}/dt)\} =
2\,(\hat{E}|\sqrt{\hat{\rho}}\,\hat{H})$ as discussed above. With
the help of $(\hat{E}_H|\sqrt{\hat{\rho}}\,\hat{H}) = 0$, Eq.
(\ref{eq:no-heat_1}) reduces to
\begin{align}
    (\hat{E}_D|\sqrt{\hat{\rho}}\,\hat{H}) = -\text{Re} \sum_n
    {\<n|\hat{\rho}\,(d/dt)|n\>}\,{{\mathcal E}_n}\,,\tag{\ref{eq:no-heat_0}b}\label{eq:E_time-dpt_1}
\end{align}
where the right-hand side is non-vanishing in contrast to its
counterpart for the time-independent Hamiltonian, which vanishes.
Substituting the identity of completeness $\sum_m |m\>\<m| = \id$
into the right-hand side of (\ref{eq:E_time-dpt_1}), recognizing
that $\<n|\partial_t|n\>$ is purely imaginary as a result of
$\<n|\partial_t|n\> + \<\partial_ t n|n\> = 0$, and then applying
the relation of instantaneous eigenstates given by \cite{GON13}
\begin{equation}\label{eq:time-dpt_2}
    \<m|\partial_t|n\> = \<m|\{\partial_t \hat{H}(t)\}|n\>/({\mathcal E}_n - {\mathcal
    E}_m)
\end{equation}
which is valid for $m \ne n$, one finally obtains the exact
expression
\begin{align}
(\hat{E}_D|\sqrt{\hat{\rho}}\,\hat{H})\; =\; -\text{Re}
\sum_n\sum_{m\,(\ne n)} \rho_{nm}\,\<m|\{\partial_t
\hat{H}(t)\}|n\>/(1 - {\mathcal E}_m/{\mathcal
    E}_n)\tag{\ref{eq:no-heat_0}c}\,.\label{eq:no-heat_2}
\end{align}
For simplicity, it is assumed here that the system is non-degenerate
(${\mathcal E}_n \ne {\mathcal E}_m$ if $n \ne m$). Eq.
(\ref{eq:no-heat_2}) can then be rewritten in terms of the
commutator $[\;,\;]_-$ as
\begin{align}
(\hat{E}_D|\sqrt{\hat{\rho}}\,\hat{H})\; =\; \sum_n\sum_{m\,(\ne n)}
\frac{\rho_{nm}}{2\,({\mathcal E}_n - {\mathcal E}_m)}\;
\<m|\left[\hat{H}, \{\partial_t
\hat{H}(t)\}\right]_-|n\>\,.\tag{\ref{eq:no-heat_0}d}\label{eq:time-dpt_2-0}
\end{align}
It should be noted that Eq. (\ref{eq:time-dpt_2-0}) can
straightforwardly be generalized to a system with a continuous
energy spectrum \cite{ILK17}. Furthermore, the validity of
(\ref{eq:time-dpt_2-0}) can easily be verified from the previous
example for $\hat{H}_0(t)$ in such a way that the left-hand side of
(\ref{eq:time-dpt_2-0}) is explicitly given by
\begin{subequations}
    \begin{equation}\label{eq:left-hand-side-1}
    (\hat{E}_D|\sqrt{\hat{\rho}}\,\hat{H}) = \frac{1}{2} \sum_n \<n|(d\hat{\rho}/dt)|n\>\; {\mathcal E}_n
    = \frac{\hbar \omega}{2}\,\{\partial_t\,\rho_{11} + a(t)\,\text{Re}(\partial_t\,\rho_{12})\}\,
    \end{equation}
and the right-hand side becomes
\begin{equation}\label{eq:time-dpt_2-1}
\frac{1/2}{{\mathcal E}_1 - {\mathcal
E}_2}\,\left\{\rho_{12}\,\<2|\left[\hat{H}, \{\partial_t
\hat{H}(t)\}\right]_-|1\> - \rho_{21}\,\<1|\left[\hat{H},
\{\partial_t \hat{H}(t)\}\right]_-|2\>\right\}
\end{equation}
\end{subequations}
which can immediately be reduced to $\hbar \omega \dot{a}\,\{4\,(1 +
a^2)\}^{-1}\,\{2 a\,\rho_{11} - a + 2a^2\,\text{Re}(\rho_{12})\}$
where $\rho_{12} = \<1|\hat{\rho}|2\>$ with $|1\>$ and $|2\>$
explicitly given by (\ref{eq:eigenenergy1})-(\ref{eq:eigenenergy2}).
The equality of this last expression with the right-hand-side of Eq.
(\ref{eq:left-hand-side-1}) confirms that $\delta Q_{\text{in}} = 0$
in (\ref{eq:heat-explicit-H_0}), which is consistent with the
assumption of no heat transfer for this system.

For purposes of the development below, Eq. (\ref{eq:no-heat_2}) is
now rewritten by first noting that Eq. (\ref{eq:density
-op-conserved_1}) is also valid for a generic time-dependent
Hamiltonian. With the help of (\ref{eq:time-dpt-inner-products_1}),
this immediately yields that $(\hat{E}_D|\sqrt{\hat{\rho}})= 0$.
Therefore, a real functional $c(\hat{\rho})$ can be introduced such
that $(\hat{E}_D|\sqrt{\hat{\rho}}\,\hat{H}) =
(\hat{E}_D|\sqrt{\hat{\rho}}\,\{\hat{H} + c(\hat{\rho})\,\id\})$.
Consistent with the case for a time-independent Hamiltonian, the
real functional $c(\hat{\rho})$ is set equal to $-{\mathcal U}(t)$.
Two normalized operators are introduced next such that $\hat{z}
:=\hat{E}_D/\{(\hat{E}_D|\hat{E}_D)\}^{1/2}$ with $(\hat{z}|\hat{z})
= 1$ and $\hat{h} := \sqrt{\hat{\rho}}\,(\Delta
\hat{H})/\sigma_{\scriptscriptstyle H}$ with $(\hat{h}|\hat{h}) = 1$
where the standard deviation $\sigma_{\scriptscriptstyle H}(t) =
\{(\sqrt{\hat{\rho}}\,\Delta\hat{H}|\sqrt{\hat{\rho}}\,\Delta\hat{H})\}^{1/2}$
as in the time-independent Hamiltonian case. Then,
$(\hat{E}_D|\sqrt{\hat{\rho}}\,\Delta \hat{H}) =
\{(\hat{E}_D|\hat{E}_D)\}^{1/2}\,\sigma_{\scriptscriptstyle
H}\,\cos(\theta_{zh})$ where $\cos \theta_{zh} = (\hat{z}|\hat{h})$.
This enables Eq. (\ref{eq:no-heat_2}) to be transformed into
\begin{equation}\label{eq:no-heat_4}
    \cos \theta_{zh} = \frac{\Lambda(t)}{\{(\hat{E}_D|\hat{E}_D)\}^{1/2}\,\sigma_{\scriptscriptstyle H}}\,,
\end{equation}
where $\Lambda(t) = \text{Re}\sum_n\sum_{m\,(\ne n)}
\rho_{nm}\,\<m|\{\partial_t\,\hat{H}(t)\}|n\>\,({\mathcal
E}_m/{\mathcal E}_n - 1)^{-1}$. This last equation can be used to
determine the direction of $\hat{E}_D$ as long as the magnitude
$\{(\hat{E}_D|\hat{E}_D)\}^{1/2}$ is known. In fact, it is seen from
this generalization to the case of a time-dependent Hamiltonian that
the time-independent Hamiltonian case exactly corresponds as
required to the special case of $\theta_{zh} = \pi/2$. Note also
that for the system $\hat{H}_0(t)$ given in
(\ref{eq:eigenenergy1})-(\ref{eq:eigenenergy2}), Eq.
(\ref{eq:density -op-conserved_1}) holds true, and
$(\sqrt{\hat{\rho}}\,\hat{H}_0|\sqrt{\hat{\rho}}\,\hat{H}_0) =
\text{Tr}[\hat{\rho}(t)\,\{\hat{H}_0(t)\}^2] = \{{\mathcal
E}_{1}(t)\}^2$ explicitly so that the variance is given by
\begin{equation}\label{eq:standard-deviation-H(0)_1}
    \{\sigma_{\scriptscriptstyle H}(t)\}^2 = (\hbar\omega/2)^2\,\left[\{a(t)\}^2 - 4\,\{\upsilon_0(t)\}^2 + 4\,\upsilon_0(t)\right]\,,
\end{equation}
where the dimensionless quantity $\upsilon_0(t) = \rho_{11}(t) +
a(t)\,\text{Re}\{\rho_{12}(t)\} = 1/2 + {\mathcal
U}_0(t)/\hbar\omega$.

Now, before exploring an explicit evaluation for
$(\hat{E}_D|\hat{E}_D)$, the quantity $(\hat{D}|\hat{D}) =
4\,\tau^2\,(\hat{E}_D|\hat{E}_D)$, which is more straightforward to
evaluate, is first considered. Then with the help of
(\ref{eq:dissipation2}),
\begin{eqnarray}\label{eq:oprator-D-1}
    \hat{D} &=& \left.\sqrt{\hat{\rho}}\,\hat{B}\,\ln(\hat{\rho})\right|_{\perp\,{\mathcal
    L}\{\sqrt{\hat{\rho}}\,\id,\,\sqrt{\hat{\rho}}\,\hat{\Upsilon}\}}\n\\
    &=& \sqrt{\hat{\rho}}\,\hat{B}\,\ln(\hat{\rho}) -
    \left\{(\sqrt{\hat{\rho}}\,\hat{B}\,\ln\hat{\rho}|\sqrt{\hat{\rho}})\,\sqrt{\hat{\rho}}\,\id\,
    +\, (\sqrt{\hat{\rho}}\,\hat{B}\,\ln{\hat{\rho}}|\sqrt{\hat{\rho}}\,\Delta\hat{\utilde{\Upsilon}})\,\sqrt{\hat{\rho}}\,\Delta\hat{\utilde{\Upsilon}}\right\}\,,
\end{eqnarray}
where the operator $\sqrt{\hat{\rho}}\,\hat{\Upsilon}(t)$ is described
in what follows. Thus, using $\Delta\hat{\utilde{\Upsilon}}(t) =
\hat{\Upsilon}(t) -
(\sqrt{\hat{\rho}}\,\hat{\Upsilon}|\sqrt{\hat{\rho}})\,\id$ with
$(\sqrt{\hat{\rho}}\,\Delta\hat{\utilde{\Upsilon}}|\sqrt{\hat{\rho}}\,\Delta\hat{\utilde{\Upsilon}})
= 1$ guarantees that the two operators $\sqrt{\hat{\rho}}\,\id$ and
$\sqrt{\hat{\rho}}\,\Delta\hat{\utilde{\Upsilon}}(t)$ are
orthonormal to each other so that
$(\sqrt{\hat{\rho}}\,\Delta\hat{\utilde{\Upsilon}}|\sqrt{\hat{\rho}}\,\id)
= 0$ at every instant of time. To visualize the behavior of
$\sqrt{\hat{\rho}}\,\hat{\Upsilon}(t)$, a three-dimensional space of
linear operators spanned by the orthonormal basis $\{\hat{x} \to
\sqrt{\hat{\rho}}\,\id;\,\hat{y} \to
\sqrt{\hat{\rho}}\,\Delta\hat{\utilde{\Upsilon}};\,\hat{z} \to
\utilde{\hat{E}}_D\}$ with $\utilde{\hat{E}}_D =
\hat{E}_D/\{({\hat{E}}_D|{\hat{E}_D})\}^{1/2} =
-\hat{D}/\{(\hat{D}|\hat{D})\}^{1/2}$ is introduced as illustrated
in Fig. \ref{fig:fig1}. This means that the operator
$\sqrt{\hat{\rho}}\,\Delta\hat{\utilde{\Upsilon}}(t)$ is chosen so
that
$(\sqrt{\hat{\rho}}\,\Delta\hat{\utilde{\Upsilon}}|\utilde{\hat{E}}_D)
= 0$. This three-dimensional space enables a linear operator to be
specified by its components $(x, y, z)$. For example, for
$\sqrt{\hat{\rho}}\,\hat{B}\,\ln\hat{\rho}$ given in
(\ref{eq:oprator-D-1}), $x =
(\sqrt{\hat{\rho}}\,\hat{B}\,\ln\hat{\rho}|\hat{x}) < 0$, $y =
(\sqrt{\hat{\rho}}\,\hat{B}\,\ln\hat{\rho}|\hat{y})$, and $z =
(\sqrt{\hat{\rho}}\,\hat{B}\,\ln\hat{\rho}|\hat{z}) < 0$. The
decomposition $\sqrt{\hat{\rho}}\,\Delta\hat{H}(t) =
(\sqrt{\hat{\rho}}\,\Delta\hat{H}|\hat{y})\,\hat{y} +
(\sqrt{\hat{\rho}}\,\Delta\hat{H}|\hat{z})\,\hat{z}$ then follows where
$\sqrt{\hat{\rho}}\,\Delta\hat{H}(t)/\sigma_{\scriptscriptstyle H}$
is represented by $\hat{h}$ in Fig. \ref{fig:fig1} and
$(\sqrt{\hat{\rho}}\,\Delta\hat{H}|\hat{z}) =
\sigma_{\scriptscriptstyle H}\,\cos\theta$ and
$(\sqrt{\hat{\rho}}\,\Delta\hat{H}|\hat{y}) = \pm
\sigma_{\scriptscriptstyle H}\,\sin\theta$.
Therefore, the angle $\theta_{zh}(t)$ given in (\ref{eq:no-heat_4})
is geometrically seen as the polar angle $\theta(t)$ of this
operator space. Furthermore, using the decomposition for $\sqrt{\hat{\rho}}\,\Delta\hat{H}(t)$ given above and the assignments for $(\hat{h},\hat{y},\hat{z})$ depicted above and in Fig. 1,
\begin{equation}\label{eq:y-operator-1}
    \hat{y} = \{\hat{h} - (\cos\theta)\,\hat{z}\}/(\pm \sin\theta),
\end{equation}
which can be interpreted as the projection of
$\sqrt{\hat{\rho}}\,\hat{\Upsilon}(t)$ onto the $y$-axis. As a
consequence, $\hat{D} \propto -\hat{z}$, which accordingly is
perpendicular to the $(xy)$-plane of this operator space as
required.

The magnitude of $\hat{D}$, which is now explicitly evaluated, easily with the help
of (\ref{eq:oprator-D-1}) reduces to
\begin{equation}\label{eq:D-magnitude}
    \{(\hat{D}|\hat{D})\}^{1/2} = \{(\sigma_{\scriptscriptstyle \ln\rho})^2 - (\sqrt{\hat{\rho}}\,[\hat{B}\,\ln\hat{\rho} -
    \eta\,\id]|\hat{y})\}^{1/2}\,,
\end{equation}
where $\eta =
(\sqrt{\hat{\rho}}\,\hat{B}\,\ln\hat{\rho}|\sqrt{\hat{\rho}}) =
-s/k_{\text{\tiny B}}$, $s = -k_{\text{\tiny
B}}\,\text{Tr}(\hat{\rho}\,\ln\hat{\rho})$, and the variance
$(\sigma_{\scriptscriptstyle \ln\rho})^2 =
\text{Tr}\{\hat{\rho}\,(\ln\hat{\rho})^2\} - \eta^2$ with
$\sigma_{\scriptscriptstyle \ln\rho} = \sigma_{\scriptscriptstyle
s}/k_{\text{\tiny B}}$. Eq. (\ref{eq:y-operator-1}) is next
substituted into (\ref{eq:D-magnitude}) and the relations
$\cos\theta = -(\hat{D}|\hat{h})/\{(\hat{D}|\hat{D})\}^{1/2}$ and
$(\hat{D}|\hat{D}) = (\sqrt{\hat{\rho}}\,\{\hat{B}\,\ln\hat{\rho} -
\eta\,\id\}|\hat{D})$ applied. After some algebraic manipulations,
the following quadratic equation in compact form is found: $\chi^2 +
2 \alpha\,(\cos\theta)\,\chi -
\{(\sin\theta)\,\sigma_{\scriptscriptstyle \ln\rho}\}^2 + \alpha^2 =
0$ where $\chi = (\hat{D}|\hat{D})^{1/2}$ and $\alpha(\theta) =
(\sqrt{\hat{\rho}}\,\{\hat{B}\,\ln\hat{\rho} - \eta\,\id\}|\hat{h})$
with $(\sigma_{\scriptscriptstyle \ln\rho})^2 \geq \alpha^2$. This
easily yields that
\begin{equation}\label{eq:chi_pm}
    \chi_{\pm}(\theta)\, =\, -\alpha\,(\cos\theta) \pm (\sin\theta)\,\{(\sigma_{\scriptscriptstyle \ln\rho})^2 -
    \alpha^2\}^{1/2}\,,
\end{equation}
where the signs $+/-$ are in accordance with their order on both
sides and $\chi_+ \geq \chi_{-}$. Substituting the two roots
$\chi_{\pm}$ into (\ref{eq:D-magnitude}), it is concluded that
$\chi_+
> 0$ is the only allowed solution consistent with the requirement
that $\{(\sin\theta)\,\sigma_{\scriptscriptstyle \ln\rho}\}^2 >
\{\alpha(\theta)\}^2$. For the case of $\theta = \pi/2$ at a given instant of time,
$\chi_+(\pi/2) = [(\sigma_{\scriptscriptstyle \ln\rho})^2 -
\{\alpha(\pi/2)\}^2]^{1/2}$, which corresponds to the case of the
time-independent Hamiltonian. In contrast, if $\theta = 0$ or
$\pi$ at a given instant of time, then $\alpha = 0$ and $\chi_+ = 0$, which corresponds to the case of
no entropy-generation.

The inner product $(\hat{E}_D|\hat{E}_D)$ is now determined by first
considering the inequality given by $\{(\hat{E}_D|\hat{E}_D)\}^{1/2}
\geq (\hat{E}_D|\hat{h}) =
(\cos\theta_{zh})\,\{(\hat{E}_D|\hat{E}_D)\}^{1/2}$. Recall that
Eqs. (\ref{eq:no-heat_1})-(\ref{eq:time-dpt_2-0}) have been obtained
directly from Eq. (\ref{eq:no-heat_0}), which is physically relevant
since it is required by thermodynamics and by the actual dynamics of
the density operator $\hat{\rho}$ of (\ref{eq:total_E_1}). In fact,
they are the only available expressions, which implicitly contain
information on the magnitude of the dynamics of the operator
$\hat{E}_D$. Motivated by this fact, an approach can now be proposed
to determine $(\hat{E}_D|\hat{E}_D)$ in such a way that without
changing the Hamiltonian $\hat{H}(t)$ and the internal energy
${\mathcal U}(t)$ at time $t$, $|(\hat{E}_D|\hat{h})|$ is maximized
by replacing $\hat{\rho}(t)$ with all possible density matrices
($\hat{\varrho}$'s) that have the same entropy $s(t) =
-k_{\text{\tiny B}}\,\text{Tr}\{\hat{\rho}(t)\,\ln\hat{\rho}(t)\}$
(or with the purity $\mu(t) = \text{Tr}\{\hat{\rho}^2(t)\}$ in a
weaker form). The maximum value $(\hat{E}_D|\hat{h})_{\mbox{\tiny
max}}$ at time $t$ is then identified as
$\{(\hat{E}_D|\hat{E}_D)\}^{1/2}$, which is subsequently substituted
into (\ref{eq:no-heat_4}) to determine the angle $\theta_{zh}(t)$.
Here it is stressed that, when the angle $\tilde{\theta}_{zh} = 0$
in this identification, it does not represent the actual angle
$\theta_{zh} = 0$ between $\hat{E}_D$ and $\hat{h}$ which simply
corresponds to the case when $\hat{E}_D = 0$ (i.e., $\chi_+ = 0$) as
discussed in the previous paragraph.

The intra-relaxation time $\tau =
\{(\hat{D}|\hat{D})\}^{1/2}/[2\,\{(\hat{E}_D|\hat{E}_D)\}^{1/2}]$
determined by this approach is more physically relevant than its
minimum-uncertainty counterpart since the former reflects the actual
dynamics of the density operator $\hat{\rho}(t)$ in terms of
$\mu(t)$, especially for a mixed state $\hat{\rho}(t)$ with $\mu(t)
< 1$. The detailed development for $\tau$ is given below. Therefore,
this value of the relaxation time is necessarily greater than the
minimum-uncertainty value corresponding to a pure state only (more
precisely, to the (instantaneous) ground state of the system
considered). The latter time is completely irrelevant to the actual
dynamics. As a consequence, it is argued here that the maximizing
process proposed above for determining the {\em magnitude} and {\em
direction} of $\hat{E}_D$ (and, thus, the magnitude of $\tau$) must
be regarded as an important addition to the SEAQT framework, one not
considered thus far even for the case of a time-independent
Hamiltonian.

The inner product $(\hat{E}_D|\hat{E}_D)$ is now determined for the example
$\hat{H}_0(t)$ previously used. With the help of
(\ref{eq:left-hand-side-1}) and
(\ref{eq:standard-deviation-H(0)_1}), it is straightforward to
obtain
\begin{equation}\label{eq:time-inner-product-2}
    (\hat{E}_D|\hat{h}) = \frac{\partial_t\,\upsilon_0(t) -
    \{\text{Re}(\rho_{12})\}\,\partial_t\,a(t)}{[\{a(t)\}^2 - 4\,\{\upsilon_0(t)\}^2 + 4\,\upsilon_0(t)]^{1/2}}\,.
\end{equation}
For a fixed purity $\mu_0(t) = (\rho_{11})^2 + (\rho_{22})^2 +
2\,|\rho_{12}|^2$ at time $t$, the right-hand side of
(\ref{eq:time-inner-product-2}) is maximized by finding an optimal
value of $\text{Re}(\varrho_{12})$ to replace
$\text{Re}\{\rho_{12}(t)\}$. To do so, the maximum value
$\{\text{Re}(\rho_{12})\}^2_{\mbox{\tiny max}}$, which minimizes
$(\rho_{11})^2 + (\rho_{22})^2$ in the purity measure, is found,
resulting in $\rho_{11} = \rho_{22} = 1/2$. It then follows that
$\{\text{Re}(\rho_{12})\}^2_{\mbox{\tiny max}} = [2\,\mu_0(t) - 1 -
4\,\{\text{Im}(\rho_{12})\}^2]/4$. The maximum of this maximum,
$\{\text{Re}(\rho_{12})\}^2_{\mbox{\tiny max,max}}$, occurs with
$\text{Im}(\rho_{12}) = 0$. $\{\text{Re}(\varrho_{12})\}^2 =
\{\text{Re}(\rho_{12})\}^2_{\mbox{\tiny max,max}}$ is then
substituted for $\text{Re}(\rho_{12})$ in
(\ref{eq:time-inner-product-2}) and the inequality $r_1 + r_2 \leq
|r_1| + |r_2|$ used for two real numbers $r_1$ and $r_2$ to arrive
at
\begin{equation}\label{eq:time-inner-products-3}
    (\hat{E}_D|\hat{E}_D)^{1/2} =
    \frac{|\partial_t\,\upsilon_0(t)| + |\{\text{Re}(\varrho_{12})\}\,
    \partial_t\,a(t)|}{[\{a(t)\}^2 - 4\,\{\upsilon_0(t)\}^2 + 4\,\upsilon_0(t)]^{1/2}}
\end{equation}
where the two constraints on $\upsilon_0$, i.e., ${\mathcal U}_0$
with $a(t)$, and $\mu_0$ hold. By substituting
(\ref{eq:time-inner-products-3}) into (\ref{eq:no-heat_4}), the direction of $\hat{E}_D$ denoted by
$(\theta_{zh})_0$ can be determined.

Based on the above analysis, the
internal-relaxation time can be uniquely determined. Using (\ref{eq:no-heat_4}) in (\ref{eq:chi_pm}) results in
\begin{equation}\label{eq:no-heat_7-1}
    \tau(\hat{\rho}) = \frac{\chi_+}{2\,\{(\hat{E}_D|\hat{E}_D)\}^{1/2}} =
    \frac{-\alpha\,\Lambda + (\chi_{+,0})\,\{(\hat{E}_D|\hat{E}_D)\, \sigma_{\scriptscriptstyle H}^2 -
    \Lambda^2\}^{1/2}}{2\, \sigma_{\scriptscriptstyle H}\,
    (\hat{E}_D|\hat{E}_D)}\,,
\end{equation}
where $\chi_{+,0}(\theta_{zh}) = [(\sigma_{\scriptscriptstyle
\ln\rho})^2 - \{\alpha(\theta_{zh})\}^2]^{1/2}$ and
$(\hat{E}_D|\hat{E}_D)$ is found from the maximization process described above. Therefore, all quantities on the right-hand side of (\ref{eq:chi_pm}) can be evaluated. Obviously, Eq. (\ref{eq:no-heat_7-1}) is also valid
for the special case of a time-independent Hamiltonian for which $\theta_{zh} = \pi/2$ and $\Lambda = 0$, leading to $\tau \to
\{\chi_+(\pi/2)\}\,[2\,\{(\hat{E}_D|\hat{E}_D)\}^{1/2}]^{-1}$, which is clearly different from its minimum-uncertainty counterpart
$\hbar\,(2\,\sigma_{\scriptscriptstyle H})^{-1}$. As seen in
(\ref{eq:no-heat_7-1}) [cf. $\Lambda(t)$], the off-diagonal terms of the density matrix play a critical role in determining
$\tau(\hat{\rho})$. In contrast, the minimum-uncertainty value
results from the ground (minimum-energy) pure state for which the off-diagonal terms are identically zero. For the more general case of $\theta_{zh} \ne \pi/2$ and $\Lambda
\ne 0$ (i.e., for the case of the time-dependent Hamiltonian) and with the
help of (\ref{eq:no-heat_4}), the expression for $\tau$ can be rewritten as
\begin{equation}\label{eq:no-heat_7-0}
    \tau(\hat{\rho}) = -(\hat{D}|\sqrt{\hat{\rho}}\,\Delta\hat{H})\, \{2\, \Lambda(t)\}^{-1}\,,
\end{equation}
where $(\hat{D}|\sqrt{\hat{\rho}}\,\Delta\hat{H}) =
-\{(\hat{D}|\hat{D})\}^{1/2}\,\sigma_{\scriptscriptstyle
H}\,(\cos\theta_{zh})$.

%
\section{Conclusions}\label{sec:sec5}
%
The preceding development, which is based on a formal consideration of time-dependent Hamiltonians ($\hat{H}(t)$'s), is a generalization of the SEAQT framework that results in \textit{ab initio} expressions for the intra-relaxation time. The latter is an important element of this framework, one which had not previously been uniquely determined as an integral part of the theory. The approach proposed here to determine $\tau{(\hat{\rho})}$
is a physically relevant one based on an additional maximization process, i.e. one that supplements the steepest-entropy-ascent maximization, which forms the basis of the SEAQT framework. The expressions developed are valid for both time-dependent and time-independent Hamiltonians and transform the description provided by this framework into an even more robust one at the
fundamental level.

The other significant development provided here is that of critically contributing to a resolution of a fundamental issue of thermodynamics in the quantum domain concerning
the unique definition of {\em process-dependent} work and heat functions. This is done with the aid of the SEAQT framework and the energy balance resulting from the first law of thermodynamics. As is well-known, this conceptual problem has been an open
question within the thermodynamics embedded in the standard quantum mechanics approach when both work, as given by an explicitly time-dependent Hamiltonian, and heat are simultaneously considered. It is this latter development, which will be a particular focus of a future paper. An additional focus will be the
numerical application of our framework to a number of driven quantum systems such as the two-level system with $\hat{H}_0(t)$
introduced in Section \ref{sec:sec3} and a linear oscillator with a time-dependent frequency. These applications will take advantage of the fact that the numerical implementation of the SEAQT framework
has thus far been very robust for the case of time-independent Hamiltonians.

Finally, a consequence of the developments given here is that SEAQT is not just an alternative approach to thermodynamics in the quantum domain but in fact sheds new light on the various fundamental but not completely resolved
questions of thermodynamics. It is also expected that these new developments will contribute to providing foundational guidance for driven thermodynamic machines operating in the quantum/nano domain.

\section*{Acknowledgments}
The first author dedicates this work to the late G\"unter Mahler (Stuttgart). He also thanks Peter Salamon (San Diego) for a helpful
discussion on the unique definition of process-dependent heat and work functions in the quantum domain during the workshop
``Thermodynamics and Nonlinear Dynamics in the Information Age'', Telluride/Colorado in 2015. He gratefully acknowledges the financial support
provided by the U.S. Army Research Office (Grant No. W911NF-15-1-0145).

\newpage
\begin{figure}[htb]
\centering\hspace*{-3cm}\vspace*{-0cm}{
\includegraphics[scale=1.5]{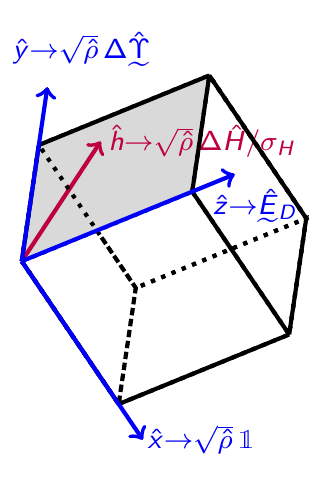}
\caption{\label{fig:fig1}}}
\end{figure}
Fig.~\ref{fig:fig1}: (Color online) The three-dimensional space of
linear operators whose basis consists of  the three orthonormal
operators $\{\hat{x} = \sqrt{\hat{\rho}}\,\id;\,\hat{y} =
\sqrt{\hat{\rho}}\,\Delta\hat{\utilde{\Upsilon}};\,\hat{z} =
\utilde{\hat{E}}_D\}$ [cf. after Eq. (\ref{eq:oprator-D-1})]. Here
the normalized operator $\hat{h} = (\sin\theta)\,\hat{y} +
(\cos\theta)\,\hat{z} =
\sqrt{\hat{\rho}}\,\Delta\hat{H}(t)/\sigma_{\scriptscriptstyle H}$
lying on the $(yz)$-plane, expressed in terms of the polar angle
$\theta$ of the spherical coordinate system $(r, \theta, \varphi)$,
where $0 \leq \theta \leq \pi$, and $\sin\theta = (\hat{h}|\hat{y})$
and $\cos\theta = (\hat{h}|\hat{z})$.

\begin{thebibliography}{1}
\bibitem{KIM17} For the case of a time-dependent Hamiltonian, one may
also consider that a given observable $\hat{A}(t)$ explicitly depends on
time. Then, following the same guidelines as applied above for the
case of $\hat{A}$ not being explicitly time-dependent (cf. \cite{GRI05}),
one can obtain $\sigma_{\scriptscriptstyle A} = |d \<\hat{A}\>/dt -
\<\partial \hat{A}/\partial t\>|\,\tilde{\tau}_{\scriptscriptstyle
A}$ where the modified characteristic time
$\tilde{\tau}_{\scriptscriptstyle A} := \tau_{\scriptscriptstyle
A}\,\{1 - |(\<\partial A/\partial t\>)/(d \<\hat{A}\>/dt)|\}^{-1}$.
It then follows that $\sigma_{\scriptscriptstyle
H}\,\tilde{\tau}_{\scriptscriptstyle A} \geq \hbar/2$. The
interpretation of $\tilde{\tau}_{\scriptscriptstyle A}$ should be
modified correspondingly from that of $\tau_{\scriptscriptstyle A}$.
The time $\tilde{\tau}_{\scriptscriptstyle H}$, corresponding to
$\tau_{\scriptscriptstyle H}$ given for the case of a
time-independent Hamiltonian and likewise chosen as the minimum
value of all possible characteristic times
$\tilde{\tau}_{\scriptscriptstyle A}$, follows immediately.
\bibitem{ILK17} In case that the energy spectrum is continuous,
we first apply the analytic continuation in such a way that the
right-hand side of (\ref{eq:time-dpt_2-0}) is modified by the
replacement of $\sum_n\sum_{m\,(\ne n)} \to \int\int dn\,dm$ and $x
:= {\mathcal E}_n - {\mathcal E}_m \to {\mathcal E}_n - {\mathcal
E}_m\,\pm\,i 0^+$. We next apply the identity $1/(x \pm i 0^+) =
\text{PV}(1/x)\,\mp\,i\pi\,\delta(x)$ to the accordingly modified
form of (\ref{eq:time-dpt_2-0}), where the symbol $\text{PV}$
denotes the principal value \cite{VLA71}. Then the additional
contribution of delta function to (\ref{eq:time-dpt_2-0}), appearing
when ${\mathcal E}_n = {\mathcal E}_m$ with $m = n$ (i.e., the
diagonal terms of $\hat{\rho}$) only, will vanish indeed, due to the
factor $\<n|\left[\hat{H}, \{\partial_t \hat{H}(t)\}\right]_-|n\>$
vanishing.
\end{thebibliography}
\end{document}